\documentclass[a4paper,12pt,fleqn]{article}
\usepackage{amssymb}
\usepackage{amsmath}
\usepackage{geometry}
\usepackage{graphicx}
\begin{document}
\begin{center}
\textbf{\Large
Relativistic Warning to Space Missions Aimed to Reach Phobos}
\footnote{Video report of this topic on the scientific seminar in PFUR:\\
\texttt{http://www.youtube.com/watch?v=VG3Z\_j97Y0Y} (in English)\\
\texttt{http://www.youtube.com/watch?v=fNqvcsYkix0} (in Russian)}
\\[1em]
\bf {Alexander P. Yefremov} \\[0.5em] \it Institute of Gravitation and Cosmology of Peoples' Friendship University of Russia
\\ E-mail: a.yefremov@rudn.ru
\end{center}
\textbf{Abstract}
\\[0.2em]
{\small Disagreement in estimations of the observed acceleration of Phobos yields several theories em-pirically modifying classical description of motion of the satellite, but its orbital positions de-tected by Mars-aimed spacecraft differ from predictions. It is shown that the satellite's orbital perturbations can be explained as manifestations of the relativistic time-delay effect ignored in classical models. So computed limits of Phobos' acceleration essentially exceed the experimental values. The satellite's expected orbital shift is calculated for the moment of contact with a land-ing module of the Phobos-Grunt project; the shift assessed in kilometers may prevent the mission success. Limits of the apparent relativistic accelerations are predicted for fast satellites of Jupiter.}
\\[0.8em]

\vskip 1.5em
\section*{\large 1. Introduction: Phobos' irregular motion, theory and practice}
A satellite of the planet Mars, Phobos, discovered by A.Hall in 1877 still attracts great attention. In 1911 after years of observations H.Struve offered a classical theory of Martian satellites' motion taking into account the planet's oblate shape and solar gravity. In 1945 B.Sharpless discovered a secular increase of Phobos velocity${}_{ }$[1] and surmised that the moon was spiraling in toward Mars. I.Shklovski ascribing the orbit's decay to atmospheric friction concluded that the moon could be hollow [2], maybe artificial, but the orbit's evolution was also referred to influence of tidal forces [3]. Later mathematical models [4-6] were developed in attempts to better explain observational results, some of conclusions though uncertain about the acceleration value and even sign.

Cosmic era made Phobos a desired but hardly accessible goal. In 1988 Russian Phobos-1 (said badly operated) passed by the target while Phobos-2 disappeared at 50 \textit{m} from the moon's surface. In 1999 Mars Climate Orbiter (NASA, also said badly operated) was lost near Phobos' orbit, and Mars Polar Lander vanished hardly touching the Martian atmosphere. In 2003 Beagle-2 (UK) shared the destiny without any firm conclusion of the loss. Survivors showed deficiency of existing theories: Mariner 9 (NASA, 1971) [7, 8] and Mars Express (ESA, 2004) [9] found Phobos in kilometers ahead of its expected position. A new space mission Phobos-Grunt (Russia) is planned soon [10]; if its computer program determines the target position using old models, the project may have problems.

Another reason of the Phobos' motion irregularity is considered here on the base of relativity theory. Section 2 comprises deduction of formulae for apparent acceleration and for relativistic shift of a solar system planet's satellite observed from the Earth, using methods of quaternion model of relativity. In Section 3 calculated and experimental values of the Phobos' acceleration are compared, and the shift value is assessed for the Phobos-Grunt space mission. A compact discussion is found in Section 4 with prediction of relativistic shift of fast satellites of Jupiter potentially observed from the Earth.

\section*{\large 2. Relativistic explanation}
Let the Earth (frame of reference $\Sigma $) and a planet of the solar system ($\Sigma '$, e.g. Mars) have circular trajectories (for simplicity) in ecliptic plane and revolve about the Sun with velocities of constant values $V_{E} ,\, \, V_{P} $. A planet's satellite with orbital period $T'={\rm const}$ (from viewpoint of $\Sigma '$) can be regarded as a clock. The value of $\Sigma-\Sigma '$ relative velocity is found as
\begin{equation}\label{GrindEQ__1_}
V=\sqrt{V_{P}^{2} +V_{E}^{2} -2\, V_{P} V_{E} \cos \Omega t} ,  
\end{equation}
\noindent where $\Omega $ is the difference of orbital angular velocities of the Earth and the planet, $\Omega t$ is a $({\bf V}_{E} ,{\bf V}_{P} )$-angle, its zero initial value is chosen at the planets' opposition point (where the Martian satellites are usually observed optically [6]). The relative velocity value is always different from zero, $V\ne 0$, hence a relativistic time-delay effect exists. A clock belonging to $\Sigma '$ should be slow in $\Sigma $, i.e. the satellite, as a point of $\Sigma '$-clock's arrow, should be seen in $\Sigma $ at earlier position on its orbit than it is in $\Sigma '$. Emphasize two features of the effect. First, it is accumulated with time since the satellite's apparent shift increases, so the effect is potentially detected. Second, the $\Sigma $-observer will find the satellite's motion non-uniform, since the relative velocity is variable, $V=V(t)$, the frames $\Sigma $, $\Sigma '$ being non-inertial. This hampers computation of the shift-effect by means of Special Relativity (SR) valid for inertial frames of reference, though SR can be applied locally as it is done in [11] in the deduction of formula for the Thomas precession. But here assessment of the relativistic shift is done with the help of a more ``technological'' approach based on quaternion square root from SR space-time interval, the method admitting computation of relativistic effects for arbitrary frames without addressing tensor calculus of general relativity. The quaternion model of relativity theory is described in detail in Ref. [12], below its very short description is given.

It is straightforwardly verified that multiplication of quaternions, the hypercomplex numbers built on one scalar (ordinary) unit and three non-commutative vector units ${\bf q}_{k} $, is invariant under rotations of the vector units by matrices belonging to special orthogonal group with complex parameters
\begin{equation} \label{GrindEQ__2_} 
\, {\bf q}_{n'} =O_{n'k} {\bf q}_{k}  
\end{equation} 
(summation in the repeating indices is assumed here and further on), $\, O_{n'k} \in SO(3,\mathbb{C})$, the group being 1:1 isomorphic to the Lorentz group. It is also proved that similar rotations keep form of the vector-quaternion
\begin{equation} \label{GrindEQ__3_} 
\, d{\bf z}=(dx_{k} +idt\, e_{k} )\, {\bf q}_{k} ,  
\end{equation} 
under the space-time orthogonality condition $\, dx_{k} e_{k} =0$, where$\, \{ dx_{k} ,\, dt\} $ are differentials of a particle's space-time coordinates in a frame $\, \Sigma \equiv \{ {\bf q}_{k} \} $,$\, e_{k} $ is any unit vector. The square of the vector-interval \eqref{GrindEQ__3_} yields $\, (d{\bf z})^{2} =dt^{2} -dr^{2} $, the Minkowski-type space-time interval of SR, so instead of invariance of this interval one can analyze form-invariance of $\, d\textbf{z}$ thus obtaining all cinematic effects of SR with an additional advantage to consider non-inertial frames of reference [12, 13]. Apply the method for computation of characteristics of the satellite's motion estimated by an Earth's observer.

The form-invariant vector-interval describing the relativistic system ``Earth-planet (satellite)'' is chosen in the form automatically satisfying the space-time orthogonality condition
\begin{equation} \label{GrindEQ__4_} 
\, d\textbf{z}=i\,cdt'{\bf q}_{1'} =i\, cdt({\bf q}_{1} +\frac{V}{c} \, {\bf q}_{2} ) 
\end{equation} 
the fundamental velocity $c$ is constant, $dt'$ is a proper time interval in $\, \Sigma '$, $\, dt$ is respective time interval of the observer. The cinematic situation described by \eqref{GrindEQ__4_} is equivalent to the $\, \Sigma -\Sigma '$ transformation of the type \eqref{GrindEQ__2_} with the matrix
\begin{equation} \label{GrindEQ__5_} 
O_{k'n} =\left(\begin{array}{ccc} {\cosh \psi } & {-i\sinh \psi } & {0} \\ {i\sinh \psi } & {\cosh \psi } & {0} \\ {0} & {0} & {1} \end{array}\right) 
\end{equation} 
what leads to standard expression for relative velocity as a function of hyperbolic parameter $V/c=\tanh \psi $, and to the time-delay relation
\begin{equation} \label{GrindEQ__6_} 
dt'=\, dt/\cosh \psi  
\end{equation} 
apparently the same as in SR but valid for the non-inertial case. Now let the time-interval $dt'\to \, T'$ be period of the satellite's revolution measured in$\, \Sigma '$ (in fact, a physically real period, in this case is small compared to $t$, time of observation), and $dt\to \, T$ be the similar ``period'' (here a variable magnitude) observed from$\, \Sigma $. Then \eqref{GrindEQ__6_} acquire the form
\begin{equation} \label{GrindEQ__7_} 
T'=T/\cosh \psi =T\, \sqrt{1-V^{2} (t)/c^{2} }  
\end{equation} 
so the period observed from the Earth is always greater that the real one $T>T'$. But search for the period's difference, whatever desirable it could be, is of no use since relativistic corrections, important as will be shown below, would be slurred over by uncertainty of our knowledge of the involved magnitudes: gravitational constant [14], the planet's and the satellite's physical parameters [15]. So only the limits of the satellite's acceleration value and an integral apparent shift will be assessed with possible accuracy.

Making future expressions compact, denote
\[A\equiv (V_{P}^{2} +V_{E}^{2} )/2c^{2} <<1, B\equiv \, V_{P} V_{E} /c^{2} <<1\] 
and using (1, 7) find (up to the small \textit{A}, \textit{B}) the difference between the values of satellite's real orbital velocity $V'_{S} $ estimated in $\Sigma '$, and observed velocity $V_{S} (t)$ estimated in $\Sigma $
\begin{equation} \label{GrindEQ__8_} 
V'_{S} -V_{S} (t)=2\pi \, r\left(\frac{1}{T'} -\frac{1}{T} \right)=V'_{S} (A-B\cos \Omega t) 
\end{equation} 
\textit{r} being the radius of the satellite's orbit. Differentiation of \eqref{GrindEQ__8_} with respect to time of observation leads to formula for apparent satellite's acceleration
\begin{equation} \label{GrindEQ__9_} 
a=\frac{dV_{S} (t)}{dt} =-V'_{S} \Omega B\sin \Omega t 
\end{equation} 
Now using \eqref{GrindEQ__8_} find the satellite's orbital shift, the difference between its real position and that observed from the Earth
\begin{equation} \label{GrindEQ__10_} 
\Delta l\equiv \int (V'_{S} -V_{S} )\, dt =V'_{S} \, (\, At-B\, \sin \Omega t\, /\Omega ) 
\end{equation} 
the integration constant is chosen zero assuming no shift at the beginning of the observation. As expected the shift's value monotonously grows linearly in time with imposed cyclic displacements having period of the oppositions.

\section*{\large 3. Computations of the effects for the relativistic Earth-Mars (Phobos) system}
Let the planet and satellite be Mars and Phobos. Observed secular acceleration of Phobos cited in literature varies from [1] $a_{\exp } =+1.88\times 10^{-3} \deg yr^{-2} $ to zero [3] and to the negative value [4, 6] $a_{\exp } =-0.83\times 10^{-3} \deg yr^{-2} $, in degrees of the satellite's longitude in a year, one degree of Phobos' orbit (1/180 of the orbit length) equal to $327\, km$. Calculation of respective relativistic values requires the following (conventional) data available in many sources, e.g. in [15]
\[
\begin{array}{ll}
\mbox{fundamental velocity}  & c=2.997\times 10^{10} \, cm\, \, s^{-1} \\
\mbox{Earth's orbital velocity} & V_{E} =2.978\times 10^{6} \, cm\, \, s^{-1}\\
\mbox{Mars' mean orbital velocity}  & V_{P} =2.413\times 10^{6} \, cm\, \, s^{-1}\\
\mbox{Earth-Mars angular velocity difference} & \Omega =0.932\times 10^{-7} \, s^{-1}\\
\mbox{Phobos' orbital velocity} & V'_{S} =2.14\times 10^{5} cm\, \, s^{-1}
\end{array}
\]
 then the small unit-free coefficients are found as $A=8.174\times 10^{-9} $, $B=7.997\times 10^{-9} $.

Using formula \eqref{GrindEQ__9_} find upper limit (amplitude) of the apparent acceleration caused by relativistic reasons
\begin{equation} \label{GrindEQ__11_} 
a_{\max } =V'_{S} \Omega \, B=1.59\times 10^{-10} cm\, \, s^{-2} =4.84\times 10^{-3} \deg yr^{-2}  
\end{equation} 
Thus the experimental acceleration values are found well inside the limits \eqref{GrindEQ__11_} of the apparent acceleration $-a_{\max } <a_{\exp } <a_{\max } $, its sign depending on the observational data obtained before or after the opposition point. In particular, if the data fixed in a month before the Mars-Earth opposition is compared with that obtained at the opposition point, then the conclusion should be made that the satellite moves with an acceleration having the value cited in Ref. [1]. Vice versa, the satellite's deceleration [4, 6] should be detected if the observation is done a couple of weeks after the opposition peak.

Now turn to Eq. \eqref{GrindEQ__10_} to assess the Phobos' apparent orbital shift for parameters of the mission Phobos-Grunt. The mission is planned to start at the end of 2011, and it is expected to reach Phobos within $1.25\, yr$ of flight\footnote{Unfortunately the mission failed at the launching stage (9 Nov. 2011), as reported due to technical reasons.}. The satellite's orbital parameters for the spacecraft's computer program could be obtained at the last opposition in Jan. 2010 (i.e. plus $1.75\, yr$ to the flight time) or, for higher accuracy, in the last perihelion opposition in Aug. 2003 (plus $8.25\, yr$), thus the time intervals between the observations and the spacecraft-moon contact are $t_{1} =3yr$ or $t_{2} =9.5yr$. So if the relativistic effect is ignored, the mission can find Phobos in $\Delta l_{1} =1.55\, km$ or $\Delta l_{2} =5.18\, km$ ahead of its expected position, as earlier missions did. These shifts appear to be not too great compared to the moon's size ($20\, km$), moreover, corrections of spacecraft's Martian orbits are foreseen. But the shift-effect seems worth to be taken into account in advance since a light signal correcting the spacecraft position will have to cover twice the Earth-Mars distance of $2.15\times 10^{8} \, km$ (at the planned contact moment), and it will do it within 24 minutes, a time sufficient for $3000\, km$ displacement of the spacecraft on its Martian orbit (recall for comparison the last $50\, m$ of Phobos-2).

An independent effect of apparent replacement of Phobos arises when distance between the satellite and an observer changes, light velocity being finite. Elliptic shape of the Mars' orbit makes this phenomenon essential for the Earth's observer; as well a Phobos' virtual acceleration must be detected by a spacecraft as it approaches the moon. But this effect is obvious and hopefully is taken into account in any space mission.

\section*{\large 4. Discussion and a prediction}
The above given formulas and numbers should of course be regarded only as a zero-iteration to a mathematical job good enough for engineering purposes. Strict computational technology must take into account a series of essential details, among them eccentricity of the planets' orbits, dependence on time of the velocities, and certainly reliable values of the satellite's dynamic parameters refined from synthesis of observations and theoretic considerations, e.g. solution of the moon's equation of motion in Schwarzschild (or even Kerr) gravity as well as gravitational influence of other moons. But realization of these improvements is technologically clear, and if necessary it can be successfully performed. Nonetheless the shift-effect is noted. In reality its existence will hardly cause troubles for spacecraft aimed to explore a planet due to tiny probability to meet a small moon. But if the goal is the moon itself the effect may become important. Hence to a certain extent it must be taken into consideration, and in particular in the planned Phobos-Grunt project; otherwise the mission will be under a noticeable ``relativistic danger''.

Note in conclusion that the relativistic shift-effect potentially can be detected in the motion of other satellites of the solar system planets; e.g. the Earth-Jupiter relative motion should cause apparent acceleration of fast satellites of Jupiter. Assess the range of the acceleration values for the fastest Jovian moons Metis and Adrastea, necessary data given below
\[
\begin{array}{ll}
\mbox{mean orbital velocity of Jupiter}& V_{P} =1.307\times 10^{6} \, cm\, \, s^{-1} \\

\mbox{Earth-Jupiter angular velocity difference}& \Omega =1.823\times 10^{-7} \, s^{-1}\\

\mbox{mean orbital velocity of Metis} &V'_{M} =3.150\times 10^{6} cm\, \, s^{-1} \\

\mbox{mean orbital velocity of Adrastea}& V'_{A} =3.138\times 10^{6} cm\, \, s^{-1}
\end{array}
\]
\noindent the unit-free coefficient being $B=4.331\times 10^{-9} $. As is done above for Phobos, the range of the acceleration values is calculated with the help of formula \eqref{GrindEQ__9_}. So it is predicted here that precise experimental measurement of parameters of Jovian satellites' motion may lead to descovery of the moons acceleration inside the limits for Metis $a_{M} \le 2.49\times 10^{-9} cm\, \, s^{-2} =5.54\times 10^{-3} \deg yr^{-2} $, one degree of Metis' orbit equal to $4,468\, km$, and for Adrastea

\noindent $a_{A} \le 2.48\times 10^{-9} cm\, \, s^{-2} =5.47\times 10^{-3} \deg yr^{-2} $, one degree of Adrastea's orbit equal to $4,503\, km$.

\end{document}